\documentclass[prb,twocolumn,showpacs]{revtex4}
\usepackage{graphicx}
\usepackage{amsmath}
\usepackage{amssymb}

\def\lsim{\
  \lower-1.2pt\vbox{\hbox{\rlap{$<$}\lower5pt\vbox{\hbox{$\sim$}}}}\ }
\def\gsim{\
  \lower-1.2pt\vbox{\hbox{\rlap{$>$}\lower5pt\vbox{\hbox{$\sim$}}}}\ }

\begin{document}

\title{Two dispersion curves for a one-dimensional interacting Bose gas \\ under zero
boundary conditions}

\author{Maksim Tomchenko}
\email{mtomchenko@bitp.kiev.ua}
\affiliation{Bogolyubov Institute for Theoretical Physics,
        14-b Metrolohichna Str., Kyiv 03680, Ukraine}

\date{\today}
\begin{abstract}
The influence of boundaries and non-point character of interatomic interaction
on the dispersion law has been studied for a uniform Bose gas in a
one-dimensional vessel. The non-point character of interaction was taken into
account using the Gross equation, which is more general than the Gross-Pitaevskii one.
In the framework of this approach, the well-known Bogolyubov dispersion mode
$\hbar\omega(k)\approx\sqrt{\left(  \frac{\hbar^{2}k^{2}}{2m}\right)
^{2}+qn\nu(k)\frac{\hbar^{2}k^{2}}{m}}$ ($q=1$) was obtained, as well as a new
one, which is described by the same formula, but with $q\approx1/2$. The new
mode emerges owing to the account of boundaries and the non-point character of
interaction: this mode is absent when either the Gross equation for a
cyclic system or the Gross-Pitaevskii equation for a cyclic system or a system
with boundaries is solved.
Capabilities for the new mode to be observed are discussed.

       \end{abstract}


\maketitle


\section{Introduction}

The dispersion law for the one-dimensional (1D) interacting Bose gas has been
calculated in a number of works. In particular, exact microscopic solutions
were obtained for the uniform gas \cite{girardeau1960,lieb1963,gaudin} and
the lower levels were found for the gas in a set of elongated traps
\cite{stringari2002,santos2003,superTG}. The experimental
\cite{esslinger2003,nagerl2009} ratio between the lowest compressional mode
and the dipole oscillation frequencies, $\omega_{B}^{2}/\omega_{D}^{2}$,
approximately agrees with the theoretical one
\cite{stringari2002,santos2003,superTG}.
In the mentioned models
\cite{girardeau1960,lieb1963,gaudin,superTG}, the actual non-point interatomic
potential $U(|\mathbf{r}_{1}-\mathbf{r}_{2}|)$ was replaced by the
point-like one, $U_{0}\delta(\mathbf{r}_{1}-\mathbf{r}_{2})$. In works
\cite{stringari2002,santos2003}, the hydrodynamic equations were solved. They
can be regarded as a consequence of the Gross-Pitaevskii (GP) equation
\cite{pit1961,gross1963}%
\begin{align}
&  i\hbar\frac{\partial\Psi(\mathbf{r},t)}{\partial t}=\left(  -\frac
{\hbar^{2}}{2m}\triangle+V_{ext}(\mathbf{r})\right.  \nonumber\\
&  +\left.  \frac{4\pi\hbar^{2}a}{m}|\Psi(\mathbf{r},t)|^{2}\right)
\Psi(\mathbf{r},t).\label{3}%
\end{align}
In the absence of external field ($V_{ext}=0$), the GP equation can be derived
from a more general Gross equation \cite{gross1957}
\begin{align}
&  i\hbar\frac{\partial\Psi(\mathbf{r},t)}{\partial t}=-\frac{\hbar^{2}}%
{2m}\triangle\Psi(\mathbf{r},t)\nonumber\\
&  +\Psi(\mathbf{r},t)\int d\mathbf{r}^{\prime}U(|\mathbf{r}-\mathbf{r}%
^{\prime}|)|\Psi(\mathbf{r}^{\prime},t)|^{2},\label{4}%
\end{align}
if the substitution $U(|\mathbf{r}_{1}-\mathbf{r}_{2}|)\rightarrow\frac
{4\pi\hbar^{2}a}{m}\delta(\mathbf{r}_{1}-\mathbf{r}_{2})$ is made. The
replacement of the actual non-point potential by the point-like one is
believed to be quite justified if the $s$-wave scattering length $a$ is much
shorter than the average interatomic distance. Theoretical predictions made
owing to this replacement approximately agree with experimental data obtained
for \textit{non-uniform} gases in a trap.

The dispersion law for the uniform 1D gas with the point interaction is
identical at the zero \cite{gaudin} and periodic \cite{lieb1963} boundary
conditions (BCs). However, as was found in the recent publication
\cite{zero-liquid}, the dispersion law in the uniform gas is different at zero
and periodic BCs for a non-point potential of the general form $U(|\mathbf{r}%
_{1}-\mathbf{r}_{2}|)$: this is the well-known Bogolyubov law
\cite{bog1947,bz1955}
\begin{equation}
E_{b}(k)=\sqrt{\left(  \frac{\hbar^{2}k^{2}}{2m}\right)  ^{2}+n_{0}\nu
(k)\frac{\hbar^{2}k^{2}}{m}}\label{2}%
\end{equation}
under periodic BCs, and the dispersion law
\begin{equation}
E_{2}(k)=\sqrt{\left(  \frac{\hbar^{2}k^{2}}{2m}\right)  ^{2}+qn_{0}%
\nu(k)\frac{\hbar^{2}k^{2}}{m}},\ q=2^{-f},\label{5}%
\end{equation}
under zero BCs. The difference consists in the factor $q=2^{-f}$, where
$f$ is the number of non-cyclic coordinates. This result is the first evidence
that boundaries strongly affect the dispersion law in a uniform system of
interacting bosons at $T\rightarrow0$. The transition to the thermodynamic
limit in this system is incorrect, because it leads to different results
for closed and open systems. This is not a trivial effect; the discussion why
it is possible can be found in work \cite{zero-liquid}. To verify this
strange, at first sight, result, we tried \cite{zero-gas2} to analyze the
problem in the framework of GP approach. It turned out that, if the Gross
equation (\ref{4}) instead of the GP one is solved, two dispersion laws are
obtained for the uniform 1D gas with the
boundaries and without an external field: Bogolyubov (\ref{2}) and new (\ref{5}) with
$q\approx1/2$. However, BCs were taken into consideration in
Ref.~\onlinecite{zero-gas2} only partially (it was taken into account that the wave
function (WF) of the system changes its behavior at the boundary, but the WF
value at the boundary was not set), and only a particular rather than general
solution was found for the WF. Below, we will solve the problem more
precisely: a general solution for the WF will be derived, and the solutions
for the WF and the dispersion law \textit{at zero BCs} will be determined.

As has already been marked \cite{zero-liquid,zero-gas2}, solution (\ref{5})
has not been found earlier because either (i) periodic BCs were used or (ii) zero BCs
were adopted, but the actual, non-point potential was replaced by the
point-like one. In the latter case, the effect becomes lost \cite{zero-liquid,
zero-gas2}.

We confine the analysis to the simplest case of uniform 1D gas. It is not easy to solve even
this problem. Some considerations concerning the
non-uniform gas in a trap are discussed in section~\ref{sec7}. Solutions are
found in sections \ref{sec2} to \ref{sec4}, and their structure is analyzed in
section \ref{sec5}. In section \ref{sec7}, the relation of the theory to the
experiment is discussed.

\section{Basic equations}

\label{sec2}

In the ideal 1D Bose gas in a trap, the condensate exists at the finite number
of particles $N$ and low $T$ (see \cite{druten1997}). In the 1D gas with point
interaction and in the presence of a trap, 1)~there is no condensate in the
Tonks-Girardeau regime at $N\leq10$ (see \cite{girardeau2001}), and 2)~the
condensate exists in the weak interaction regime at low $T$ (see \cite{shlyap2004}).
For the \textit{uniform} 1D interacting gas in a vessel (without the trap
field), the picture is as follows. Semiclassic estimations \cite{zero-gas2}
show that the condensate is not forbidden at $T\rightarrow0$. According to the
analysis \cite{pethick2008} (Chap.~15), the condensate is not
forbidden in the classical approximation at $T\rightarrow0$, whereas in the
microscopic approach, it is forbidden even at $T=0$. But the
analysis \cite{pethick2008} is valid for a cyclic system with point
interaction. However, below we study a non-cyclic system with
non-point interaction.

Consider a uniform Bose gas in a 1D vessel with zero BCs. Interaction is
considered to be non-point of the general form. We assume that the condensate
does exist, and the Gross equation is applicable. If the condensate cannot
exist in the pure 1D case, it is possible to consider a quasi-1D geometry,
namely, a 3D system in which the motion along two dimensions is frozen. The
condensate WF reads
\begin{equation}
\Psi(x,t)=R(x,t)e^{iS(x,t)/\hbar}, \label{6}%
\end{equation}
where $R$ and $S$ are real functions. Let the system be in the
interval $x\in\lbrack-L/2,L/2]$. The zero BCs mean that%
\begin{equation}
R(x=\pm L/2,t)=0. \label{7}%
\end{equation}
In the ground state, the condensate is uniform everywhere except a very narrow
region near the walls and is described by the WF
\begin{equation}
\Psi_{0}(x,t)=R_{0}(x)e^{iS_{0}(t)/\hbar},\ S_{0}=-E_{0}t. \label{8}%
\end{equation}
For small oscillations in the system,
\begin{equation}
n\equiv R^{2}(x,t)=n_{0}(x)+\tilde{n}_{0}(x,t),\ S=-E_{0}t+s_{0}(x,t).
\label{9}%
\end{equation}
Substituting those formulas into Gross equation (\ref{4}) and neglecting the
nonuniformity of $R_{0}(x)=\sqrt{n_{0}(x)}$ near the walls, we obtain, in the
linear approximation, the following equations \cite{gross1963,zero-gas2}:
\begin{equation}
\frac{\partial\tilde{n}_{0}}{\partial t}=-\frac{n_{0}}{m}\nabla^{2}s_{0},
\label{10}%
\end{equation}%
\begin{equation}
-\frac{\partial s_{0}}{\partial t}=-\frac{\hbar^{2}}{4mn_{0}}\nabla^{2}%
\tilde{n}_{0}+\int dx^{\prime}\tilde{n}_{0}(x^{\prime},t)U(|x-x^{\prime}|),
\label{11}%
\end{equation}
where $n_{0}=N/L$. In the presence of walls, stationary oscillations can be only
standing waves; therefore,
\begin{equation}
\tilde{n}_{0}(x,t)=\tilde{n}(x)T_{n}(t),\ s_{0}(x,t)=s(x)T_{s}(t). \label{12}%
\end{equation}
Solutions for $T_{n}$ and $T_{s}$ are as follows \cite{zero-gas2}:%
\begin{equation}
T_{n}(t)=\cos{\omega t},\ T_{s}(t)=\sin{\omega t}. \label{13}%
\end{equation}
Then, the equations for $\tilde{n}(x)$ and $s(x)$ take the form%
\begin{equation}
\omega\tilde{n}(x)=\frac{n_{0}}{m}\nabla^{2}s(x), \label{14}%
\end{equation}%
\begin{equation}
-\omega s(x)=-\frac{\hbar^{2}}{4mn_{0}}\nabla^{2}\tilde{n}(x)+\int
\limits_{-L/2}^{L/2}dx^{\prime}\tilde{n}(x^{\prime})U(|x-x^{\prime}|)
\label{15}%
\end{equation}
with the boundary conditions
\begin{equation}
R_{0}(x=\pm L/2)=0,\ \tilde{n}(x=\pm L/2)=0. \label{16}%
\end{equation}
Since we neglected the nonuniform character of $R_{0}(x)$ near the walls, the
equation for the ground state becomes independent, and it will not be
considered below.

Our task consists in finding \textquotedblleft elementary\textquotedblright%
\ (with the minimum number of harmonics) solutions of Eqs.~(\ref{14}%
)--(\ref{16}) and the corresponding dispersion laws $\omega(k)$. The simplest
solution of Gross equation (\ref{4}) with cyclic BCs and GP equation (\ref{3})
with zero (or cyclic) BCs is, in the linear approximation (\ref{10}),
 (\ref{11}), a single harmonic%
\begin{equation}
\tilde{n}(x)=a\cos{kx},\ s_{0}(x)=b\cos{kx} \label{17}%
\end{equation}
with the Bogolyubov dispersion law \cite{gross1963}. However, at zero BCs, a
single harmonic is not anymore the solution of Gross equation, and a
superposition of a large number of harmonics should be considered
\cite{zero-gas2}.

Elementary solutions are tried in the following four forms:

\noindent1a)
\begin{equation}
\tilde{n}(x,k^{wp})=a_{0}(2l_{0})+2\sum\limits_{l}a_{2l}(2l_{0})\cos
{[2\pi(l+\gamma)x/L]}, \label{1a-1}%
\end{equation}%
\begin{equation}
s(x,k^{wp})=b_{0}(2l_{0})+2\sum\limits_{l}b_{2l}(2l_{0})\cos{[2\pi
(l+\gamma)x/L]}, \label{1a-2}%
\end{equation}

\noindent1b)
\begin{equation}
\tilde{n}(x,k^{wp})=a_{0}(2l_{0})+2\sum\limits_{l}a_{2l}(2l_{0})\sin
{[2\pi(l+\gamma)x/L]}, \label{1b-1}%
\end{equation}
where $l=1,2,3\ldots$, $\gamma\in\left]  -1,1\right[  $, and $k^{wp}%
=2\pi(l_{0}+\gamma)/L$ is the wave vector of the wave packet center,

\noindent2a)
\begin{align}
&  \tilde{n}(x,k^{wp})=a_{0}(l_{0})+2\sum\limits_{l}a_{2l}(l_{0})\cos
{[\pi(2l+\gamma)x/L]}\label{2a-1}\\
&  +2\sum\limits_{j}a_{2j+1}(l_{0})\cos{[\pi(2j+1+\gamma)x/L]},\nonumber
\end{align}

\noindent2b)
\begin{align}
&  \tilde{n}(x,k^{wp})=a_{0}(l_{0})+2\sum\limits_{l}a_{2l}(l_{0})\sin
{[\pi(2l+\gamma)x/L]}\label{2b-1}\\
&  +2\sum\limits_{j}a_{2j+1}(l_{0})\sin{[\pi(2j+1+\gamma)x/L]},\nonumber
\end{align}
where $\gamma\in\left]  -1,1\right]  $, $l=1,2,3,\ldots$, and
$j=0,1,2,3,\ldots$. Solutions~(2a) and (2b) correspond to wave packets
centered at $k^{wp}=\pi(l_{0}+\gamma)/L$. In Ref.~\onlinecite{zero-gas2}, solutions
with $\gamma=0$ were studied. The series for $s(x)$ in cases (1b), (2a), and
(2b) are not presented. They can be obtained from the series for $\tilde
{n}(x)$ making the substitution $a_{p}\rightarrow b_{p}$, as in
Eqs.~(\ref{1a-1}) and (\ref{1a-2}).

Solutions (1a)--(2b) \textit{generalize} solutions obtained in
Ref.~\onlinecite{zero-gas2} owing to the introduced parameter $\gamma$, which
describes the fractional part of $k^{wp}$. This parameter provides the
fulfillment of required BCs.

Below, we will see that functions (\ref{1a-1})--(\ref{2b-1}), under certain
conditions, are the solutions of Eqs.~(\ref{14})--(\ref{16}). At a fixed value
of quantum number $k^{wp}$, functions (\ref{1a-1})--(\ref{1b-1}) can be
considered as expansions in the basis set of cosine or sine functions.
Function (\ref{2a-1}) is a sum of two functions; one of then can be expanded
in the complete set of cosines $\cos{(\pi(2l+\gamma)x/L)}$, and the other in
the complete set of cosines $\cos{[\pi(2j+1+\gamma)x/L]}$. Both sets are
complete with respect to the expansion of an even function $f(x)$ determined
within the interval $x\in\lbrack-L/2,L/2]$. If $\gamma=0$, the expansion in
$\cos{(\pi(2l+\gamma)x/L)}$ is a Fourier series. In turn, function
(\ref{2b-1}) is expanded in sines; at $a_{0}(l_{0})=0,$ this is an odd function.

As will be shown below, the Bogolyubov dispersion law corresponds to solutions
(\ref{1a-1})--(\ref{1b-1}), and the new dispersion law to solutions
(\ref{2a-1}) and (\ref{2b-1}). In other words, the wave packet structure
determines the dispersion law and, consequently, is a sort of quantum number.
Solutions (\ref{2a-1}) and (\ref{2b-1}) were guessed: we cannot explain them
completely from the physical viewpoint.

Let us write down \textit{all approximations used in calculations}. i)~We considered
small oscillations and, therefore,  linearize the Gross equation. ii)~We
neglected the nonuniformity of the ground-state WF near the walls (see
justification in Ref.~\onlinecite{zero-gas2}). iii)~In the expansions, we took into
account a finite number of first summands (usually of about 100; when the
number of accounted summands was taken twice as large, the results obtained
changed insignificantly).

\section{Bogolyubov dispersion law}

\label{sec3}

Consider wave packet (1a). The function $\tilde{n}(x,k^{wp})$ satisfies zero
BCs (\ref{16}) at $\gamma=\pm1/2$ and $a_{0}=0$. Substituting Eqs.~(\ref{1a-1}%
) and (\ref{1a-2}) into Eq.~(\ref{14}) and collecting the coefficients at
independent $\cos{[2\pi(l+\gamma)x/L]}$ functions, we obtain%
\begin{equation}
b_{2l\neq0}=-\frac{\omega m}{n_{0}\tilde{k}_{2l}^{2}}a_{2l}, \label{19}%
\end{equation}
where $\tilde{k}_{2l}=2\pi(l+\gamma)/L$. \textit{It is important} that the potential
should be expanded into a proper series. The potential $U(|x_{1}-x_{2}|)$ can
be expanded in a Fourier series in several ways by considering 1)~$x_{1}$ and
$x_{2}$ separately, or 2)~$|x_{1}-x_{2}|$, or 3)~$x_{1}-x_{2}$ as the
argument. This procedure was considered in detail and with examples in
Ref.~\onlinecite{ryady}. We use the simplest Fourier series with the expansion
argument $x_{1}-x_{2}$,
\begin{align}
U(|x_{1}-x_{2}|)  &  =\sum\limits_{j=0,\pm1,\pm2,\ldots}\frac{\nu(k_{j})}%
{2L}e^{ik_{j}(x_{1}-x_{2})} \label{20} \\
&  =\frac{\nu(0)}{2L}+\sum\limits_{j=1,2,\ldots}\frac{\nu(k_{j})}{L}%
\cos{[k_{j}(x_{1}-x_{2})]}, \nonumber%
\end{align}
where $k_{j}=\pi j/L$. This series exactly reproduces the initial function at
every $x_{1}$ and $x_{2}$ within the considered interval $x_{1},x_{2}%
\in\lbrack-L/2,L/2]$. It is not worth using the standard expansion usually
applied for the thermodynamic limit, because, in the case of a system with
boundaries, it distorts the potential (see Ref.~\onlinecite{zero-liquid} and, in
detail, Ref~\cite{ryady}).

Substituting series (\ref{20}) and functions (\ref{1a-1}) and (\ref{1a-2})
into Eq.~(\ref{15}), and calculating the integral, we obtain the equations
\begin{align}
0  &  =\omega b_{0}+\sum\limits_{l=1,2,\ldots}\cos{[\pi(2l+\gamma)x/L]}
\nonumber \\ &\times  \left[
2\omega b_{2l}+\frac{\hbar^{2}\tilde{k}_{2l}^{2}}{2mn_{0}}a_{2l}+\nu(\tilde
{k}_{2l})a_{2l}\right]  +\nu(0)\sum\limits_{l=1,2,\ldots}a_{2l}c_{0}^{l}/2 \nonumber \\ &+\sum
\limits_{l,j=1,2,\ldots}a_{2l}\nu(k_{2j})c_{j}^{l}\cos{[\pi(2jx/L)]},
\label{21} %
\end{align}%
\begin{equation}
c_{j}^{l}=\frac{\sin{\pi(l+\gamma-j)}}{\pi(l+\gamma-j)}+\frac{\sin
{\pi(l+\gamma+j)}}{\pi(l+\gamma+j)},\ c_{0}^{l}=c_{j=0}^{l}. \label{23}%
\end{equation}
In turn, substituting the expansion%
\begin{equation}
\cos{[2\pi(l+\gamma)x/L]}=c_{0}^{l}/2+\sum\limits_{j=1,2,\ldots}c_{j}^{l}%
\cos{[\pi(2jx/L)]} \label{22}%
\end{equation}
into Eq.~(\ref{21}) and collecting the coefficients at independent $\cos{(2\pi
jx/L)}$ functions and the constant, we obtain the system of equations
\begin{equation}
b_{0}=-\sum\limits_{l=1,2,\ldots}\frac{a_{2l}c_{0}^{l}}{2\omega n_{0}}%
\frac{2m}{\hbar^{2}\tilde{k}_{2l}^{2}}(E_{f}^{2}(\tilde{k}_{2l},0)-\hbar
^{2}\omega^{2}), \label{24}%
\end{equation}%
\begin{equation}
\sum\limits_{l=1,2,\ldots}a_{2l}\cdot c_{j}^{l}\cdot(E_{f}^{2}(\tilde{k}%
_{2l},k_{2j})-\hbar^{2}\omega^{2})\frac{2m}{\hbar^{2}\tilde{k}_{2l}^{2}}%
=0,\ j=1,2,\ldots, \label{25}%
\end{equation}%
\begin{equation}
E_{f}^{2}(\tilde{k}_{2l},k_{2j})=\left(  \frac{\hbar^{2}\tilde{k}_{2l}^{2}%
}{2m}\right)  ^{2}+\frac{\hbar^{2}\tilde{k}_{2l}^{2}}{2m}n_{0}(\nu(\tilde
{k}_{2l})+\nu(k_{2j})). \label{26}%
\end{equation}
If $c_{j}^{l}$ is written down in the form%
\begin{equation}
c_{j}^{l}=\pi^{-1}\cos{\pi l}\cdot\cos{\pi j}\cdot\sin{\pi\gamma}\cdot\left(
\frac{1}{l+\gamma-j}+\frac{1}{l+\gamma+j}\right)  , \label{27}%
\end{equation}
system (\ref{25}) can be rewritten as follows:
\begin{align}
&\sum\limits_{l=1,2,\ldots}\breve{a}_{2l}\cdot(E_{f}^{2}(\tilde{k}_{2l}%
,k_{2j})-\hbar^{2}\omega^{2})\frac{2m}{\hbar^{2}\tilde{k}_{2l}^{2}} \nonumber \\
&\times \left(
\frac{1}{l+\gamma-j}+\frac{1}{l+\gamma+j}\right)  =0\quad (j=1,2,\ldots),
\label{28} %
\end{align}%
\begin{equation}
\breve{a}_{2l}=a_{2l}\cdot\cos{\pi l}. \label{29}%
\end{equation}
Expression~(\ref{28}) represents an infinite homogeneous system of equations
for the coefficients $\breve{a}_{2l}$ and the frequency $\omega$. It has a
solution if its determinant equals zero. Let us find the solution numerically.
For this purpose, we put
\begin{equation}
\hbar^{2}\omega^{2}=\left(  \frac{\hbar^{2}k_{0}^{2}}{2m}\right)  ^{2}%
+q^{eff}n_{0}\nu(k_{0})\frac{\hbar^{2}k_{0}^{2}}{m}, \label{30}%
\end{equation}
where $k_{0}=2\pi j_{0}/L$ and $j_{0}$ is fixed. The parameter $q^{eff}$ is
smoothly changed from $-100$ to 1000 in order to determine those values, at
which the  matrix determinant vanishes. As a result, we
obtain a sequence of \textit{solutions} $q^{eff}$ and $\omega$. We enumerate all $q^{eff}%
$-values starting from the smallest one. Then, knowing
$q^{eff}$, it is easy to determine the real $q$ in the formula
\begin{equation}
\hbar^{2}\omega^{2}(k)=\left(  \frac{\hbar^{2}k^{2}}{2m}\right)
^{2}+q(k)n_{0}\nu(k)\frac{\hbar^{2}k^{2}}{m}, \label{31}%
\end{equation}
making use the relation
\begin{align}
&  \left(  \frac{\hbar^{2}k_{0}^{2}}{2m}\right)  ^{2}+q^{eff}n_{0}\nu
(k_{0})\frac{\hbar^{2}k_{0}^{2}}{m}\nonumber\\
&  =\left(  \frac{\hbar^{2}\tilde{k}^{2}}{2m}\right)  ^{2}+q(\tilde{k}%
)n_{0}\nu(\tilde{k})\frac{\hbar^{2}\tilde{k}^{2}}{m}, \label{32}%
\end{align}
in which the $l$-th $q^{eff}$-value in the sequence is associated with
$\tilde{k}_{2l}=2\pi(l+\gamma)/L$. In this way, we find $q(\tilde{k}_{2l})$.
We used $j_{0}=1$ and $25$ at the numbers of atoms $N=100$ and 1000
($L=N/n_{0}$, $n_{0}=\mathrm{const}$). Accordingly, $j,l=1,\ldots,2N$ in
Eq.~(\ref{28}), i.e. a $2N\times2N$-matrix was considered; the increase of
matrix size did not affect the results. We found that, for $\gamma=\pm1/2$ and
every $k$ from the smallest value, $(1+\gamma)2\pi/L$, to $(N+\gamma)2\pi/L$,
the relation
\begin{equation}
q(k)=1\pm0.002 \label{31b}%
\end{equation}
holds true. Thus, we found the \textit{Bogolyubov dispersion law} (\ref{31})
and (\ref{31b}).

In the numerical analysis, we used a simple potential
\begin{equation}
U(x)=\left[
\begin{array}
[c]{cc}%
U_{0}>0, & \ |x|\leq a\\
0, & \ |x|>a
\end{array}
\right.  \label{33}%
\end{equation}
with $U_{0}=0.1~\mathrm{K}$ and $a=0.1\bar{R}$, where $\bar{R}=L/N$ is the
average interatomic distance.

Equation (\ref{21}) can also be solved differently, by expanding the constant
and $\cos{(2\pi jx/L)}$ in a series of $\cos{[\pi(2l+\gamma)x/L]}$ functions,
\begin{equation}
\cos{\pi(2jx/L)}=\sum\limits_{l=0,1,2,\ldots}c_{j}^{l}\cos{[2\pi
(l+\gamma)x/L]}, \label{34}%
\end{equation}
where the term with $l=0$ is present if $\gamma=1/2$ and absent if
$\gamma=-1/2$. At $\gamma=\pm1/2$, the functions $\cos{(\pi(2l+\gamma)x/L)}$
are orthogonal and form a complete set for the expansion of even function
$f(x)$ determined within the interval $[-L/2,L/2]$. This approach results in
different equations, but they have the same solution (\ref{31}) and
(\ref{31b}) (we examined the case $\gamma=1/2$).

For wave packet~(1b), zero BC (\ref{14}) is satisfied at $\gamma=0$ and
$a_{0}=0$. This packet was considered in Ref.~\onlinecite{zero-gas2}, and its
solution is the Bogolyubov mode as well.

\section{The second dispersion law}

\label{sec4}

Let us proceed to the consideration of wave packets~(2a) and (2b). Their
structure will be discussed below in this section and in section \ref{sec5}.
It is convenient to rewrite the formulas in such a way that the both packets
could be considered simultaneously:
\begin{align}
&\tilde{n}(x,k^{wp})=\sum\limits_{l=0,\pm1,\ldots}a_{2l}e^{i\pi(2l+\gamma
(2l))x/L} \nonumber \\
&+\sum\limits_{j=0,\pm1,\ldots}a_{2j+1}e^{i\pi(2j+1+\gamma(2j+1))x/L},
\label{40}%
\end{align}%
\begin{align}
&s(x,k^{wp})=\sum\limits_{l=0,\pm1,\ldots}b_{2l}e^{i\pi(2l+\gamma(2l))x/L} \nonumber \\
&+\sum\limits_{j=0,\pm1,\ldots}b_{2j+1}e^{i\pi(2j+1+\gamma(2j+1))x/L}.
\label{41}%
\end{align}
Here,%
\begin{equation}
\gamma(p)=\left[
\begin{array}
[c]{cc}%
\gamma, & \ p>0\\
0, & \ p=0\\
-\gamma, & \ p<0,
\end{array}
\right.  , \label{42}%
\end{equation}
and, for all $p\neq0$, the equality
\begin{equation}
a_{-p}=za_{p} \label{43}%
\end{equation}
is obeyed, where $z=1$ (for all $p$'s) or $-1$ (also for all $p$'s). From
Eqs.~(\ref{40})--(\ref{43}), if $z=1$, we obtain packet~(2a) (see
Eq.~(\ref{2a-1})), and, if $z=-1$, packet ~(2b) (see Eq.~(\ref{2b-1}))
multiplied by the imaginary unit (the latter can be easily eliminated assuming
all $a_{p}$'s to be imaginary).

So, we proceed from Eqs.~(\ref{40})--(\ref{43}) with $z=1$ or $-1$.
Substituting expressions (\ref{40}) and (\ref{41}) into (\ref{14}) and
collecting the coefficients at the exponential functions $e^{i\pi
(2l+\gamma(2l))x/L}$ and $e^{i\pi(2j+1+\gamma(2j+1))x/L}$, we obtain the
equations%
\begin{equation}
a_{0}=0,\quad b_{p\neq0}=-\frac{\omega m}{n_{0}\tilde{k}_{p}^{2}}%
a_{p},\label{44}%
\end{equation}
where, $\tilde{k}_{p}=\pi(p+\gamma(p))/L$, $p=\pm1,\pm2,\ldots$.\ In order to
obtain all possible values for the wave vector $k^{wp}=\pi(l_{0}+\gamma)/L$ of
packet center, it is necessary to put $\gamma\in\lbrack-1,1]$. Additionally,
we assume that $\gamma\neq0$. Then, the denominators in the formulas presented
below differ from zero.

\begin{figure}[ht]
\centerline{\includegraphics[width=85mm]{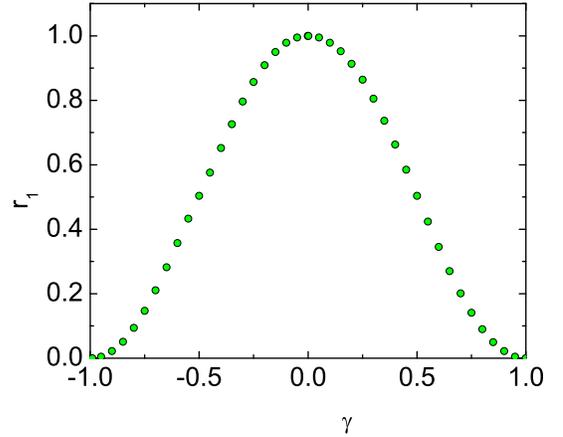}
}
\caption{Values of $r_{1}$ found for various $\gamma$'s in such a way that the
right hand side of Eq.~(\ref{47}) reproduces the left hand one the most
precisely. The maximum $|l_{1}|$ and $|j_{1}|$ were selected to equal $2N=200$.
  \label{fig1}}
\end{figure}

Substituting Eqs.~(\ref{40}) and (\ref{41}) into Eq.~(\ref{15}) and
calculating the integral, we arrive at the equation
\begin{align}
&  -\omega b_{0}-\omega\sum\limits_{l\neq0}b_{2l}e^{i\pi(2l+\gamma
(2l))x/L} \nonumber\\
& -\omega\sum\limits_{j=0,\pm1,\ldots}b_{2j+1}e^{i\pi(2j+1+\gamma
(2j+1))x/L} \nonumber\\
&  =\frac{\hbar^{2}}{4mn_{0}}  \sum\limits_{l\neq0}a_{2l}\tilde{k}%
_{2l}^{2}e^{i\pi(2l+\gamma(2l))x/L} \nonumber\\
&+\frac{\hbar^{2}}{4mn_{0}}\sum\limits_{j=0,\pm1,\ldots}%
a_{2j+1}\tilde{k}_{2j+1}^{2}e^{i\pi(2j+1+\gamma(2j+1))x/L} \nonumber\\
&  +\frac{1}{2}\sum\limits_{ll_{1}}^{l\neq0}a_{2l}\nu(k_{2l_{1}})g_{2l}%
(2l_{1}-2l)e^{i\pi2l_{1}x/L} \nonumber\\ &+\frac{1}{2}\sum\limits_{jl_{1}}a_{2j+1}%
\nu(k_{2l_{1}})g_{2j+1}(2l_{1}-2j-1)e^{i\pi2l_{1}x/L}\nonumber\\
&  +\frac{1}{2}\sum\limits_{lj_{1}}^{l\neq0}a_{2l}\nu(k_{2j_{1}+1}%
)g_{2l}(2j_{1}+1-2l)e^{i\pi(2j_{1}+1)x/L}\nonumber\\
&  +\frac{1}{2}\sum\limits_{jj_{1}}a_{2j+1}\nu(k_{2j_{1}+1})g_{2j+1}%
(2j_{1}-2j)e^{i\pi(2j_{1}+1)x/L}, \label{45}%
\end{align}
where
\begin{equation}
g_{p}(q)=\frac{2}{\pi}\frac{\sin{[\pi(q-\gamma(p))/2]}}{q-\gamma(p)}.
\label{46}%
\end{equation}
Some exponential functions in Eq.~(\ref{45}) contain $\gamma$ in their
exponents, whereas the others do not. Let us transform the exponential
functions to the identical form with the help of expansion%
\begin{align}
&e^{i\pi\gamma(p)x/L}=r_{1}\sum\limits_{l_{1}=0,\pm1,\ldots}g_{p}%
(2l_{1})e^{i2\pi l_{1}x/L} \nonumber \\
&+r_{2}\sum\limits_{j_{1}=0,\pm1,\ldots}g_{p}%
(2j_{1}+1)e^{i\pi(2j_{1}+1)x/L}, \label{47}%
\end{align}%
\begin{equation}
r_{1}+r_{2}=1. \label{49}%
\end{equation}
The exponential function is presented as a sum of two terms. The first term is
expanded in a series of \textquotedblleft even\textquotedblright\ exponential
functions $e^{i2\pi l_{1}x/L}$ (this is a Fourier series), and the second one
in a series of \textquotedblleft odd\textquotedblright\ exponential functions
$e^{i\pi(2j_{1}+1)x/L}$, which also form a complete set of orthogonal
functions. We take into account a finite number of terms in the sums in
Eq.~(\ref{47}), then the right hand side of Eq.~(\ref{47}) does not reproduce the
left hand one exactly. But the value of $r_{1}$ for various $\gamma$'s can be
selected so (see Fig.~\ref{fig1}) as to minimize the difference between the
right and left hand sides of Eq.~(\ref{47}). Below, we solve the equation for
a $200\times200$-matrix numerically. For this matrix to be obtained, it is
desirable to put the maximum $|l_{1}|$- and $|j_{1}|$-values in the sums in
Eq.~(\ref{47}) equal to 200.

Then, Eq.~(\ref{45}) contains only \textquotedblleft even\textquotedblright%
\ or \textquotedblleft odd\textquotedblright\ exponential functions and the
constant. Collecting the coefficients at each of those functions, including
the constant, equating them to zero, and making some transformations, we
obtain the equations%
\begin{align}
&b_{0}=-\sum\limits_{p=1,2,\ldots}\frac{a_{p}[g_{p}(-p)+zg_{-p}(p)]}{2\omega
n_{0}}\frac{2m}{\hbar^{2}\tilde{k}_{p}^{2}} \nonumber \\
&\times[\tilde{E}_{f}^{2}(\tilde{k}%
_{p},0,\zeta(0,p))-\zeta(0,p)\hbar^{2}\omega^{2}], \label{50}%
\end{align}%
\begin{align}
&\sum\limits_{p=1,2,\ldots}a_{p}[g_{p}(j-p)+zg_{-p}(j+p)] \label{51} \\
&\times [\tilde{E}_{f}%
^{2}(\tilde{k}_{p},k_{j},\zeta(j,p))-\zeta(j,p)\hbar^{2}\omega^{2}%
]2m/(\hbar^{2}\tilde{k}_{p}^{2})=0, \nonumber%
\end{align}%
\begin{equation}
\tilde{E}_{f}^{2}(\tilde{k}_{p},k_{j},\zeta)=\zeta\left(  \frac{\hbar
^{2}\tilde{k}_{p}^{2}}{2m}\right)  ^{2}+\frac{\hbar^{2}\tilde{k}_{p}^{2}}%
{2m}n_{0}\nu(k_{j}). \label{52}%
\end{equation}
Expression (\ref{51}) is a system of equations enumerated by the index
$j=\pm1,\pm2,\ldots$. Equations (\ref{50}) and (\ref{51}) take account of
symmetry relation (\ref{43}). If both $j$ and $p$ are either even or odd,
then, $\zeta(j,p)=r_{1}$; otherwise, $\zeta(j,p)=r_{2}$. If the sign of $j$ in
Eq.~(\ref{51}) changes, either the equation does not change or the sign before
the whole equation changes. Therefore, we consider only positive $j$'s.

\textit{It is important} that, while deriving Eqs.~(\ref{50})--(\ref{52}), we
collected coefficients before the functions $e^{i2\pi l_{1}x/L}$ and
$e^{i\pi(2j_{1}+1)x/L}$ regarded as independent. In fact, they are
dependent, but, if the exponential functions $e^{i2\pi l_{1}x/L}$ are
expanded in series of $e^{i\pi(2j_{1}+1)x/L}$ or vice versa, the matter is
expectedly reduced to a single complete set of functions, and the Bogolyubov
dispersion law is obtained for both wave packets (1a) and (1b). However, we
may collect coefficients before $e^{i2\pi l_{1}x/L}$ and $e^{i\pi
(2j_{1}+1)x/L}$ independently, without expanding either of those functions in
the set of the others. If we succeed in zeroing all the coefficients before
those functions in Eqs.~(\ref{14}) and (\ref{15}), the latter will evidently
be satisfied, i.e. we will find their solution. The solution of the problem by
expanding the function in a complete basis set is a kind of stereotype, and
its application is not useful in our case. \textit{Since the harmonics are
entangled in the integrand of Eq.~(\ref{15}), there emerges a harmonic
interplay, which results in that different dispersion laws correspond to wave
packets with different structures.}

Boundary condition (\ref{16}) bring about the equation%
\begin{equation}
(\tan{\pi\gamma/2})^{z}=z\frac{\sum\limits_{l=1,2,\ldots}a_{2l}\cos{\pi l}%
}{\sum\limits_{j=0,1,2,\ldots}a_{2j+1}\sin{\pi(j+1/2)}}. \label{53}%
\end{equation}
Thus, the coupled equations (\ref{51})--(\ref{53}) are to be solved. This can
be done in the following manner. System (\ref{51}) with an initial $\gamma
$-value is solved first. Then,  the
\textquotedblleft theoretical\textquotedblright\ $\gamma$ is determined from
the left hand side of Eq.~(\ref{53}). The
initial $\gamma$ is varied from $-1$ to 1, excluding the points $\gamma=-1$
and 0. Those initial $\gamma$'s that coincide with their theoretical
counterparts are the sought solutions.

The process is as follows. For a given initial $\gamma$-value, system
(\ref{51}) is solved and the full set of characteristic frequencies is
determined. The frequencies are enumerated, starting from the smallest one and
ascribing them the numbers $1,2,3,\ldots$. Then, some frequencies are selected
(we examined the frequencies with the numbers 1, 2, 8, 9, 29, 30, 49, 50, 99,
and 100), and the initial value of $\gamma$ is smoothly varied from $-1$ to 1
for each of them, thus finding solutions for $\gamma$. It turned out that
there are several $\gamma$-solutions for the frequency with the given number.

\begin{figure}
\includegraphics[width=0.47\textwidth]{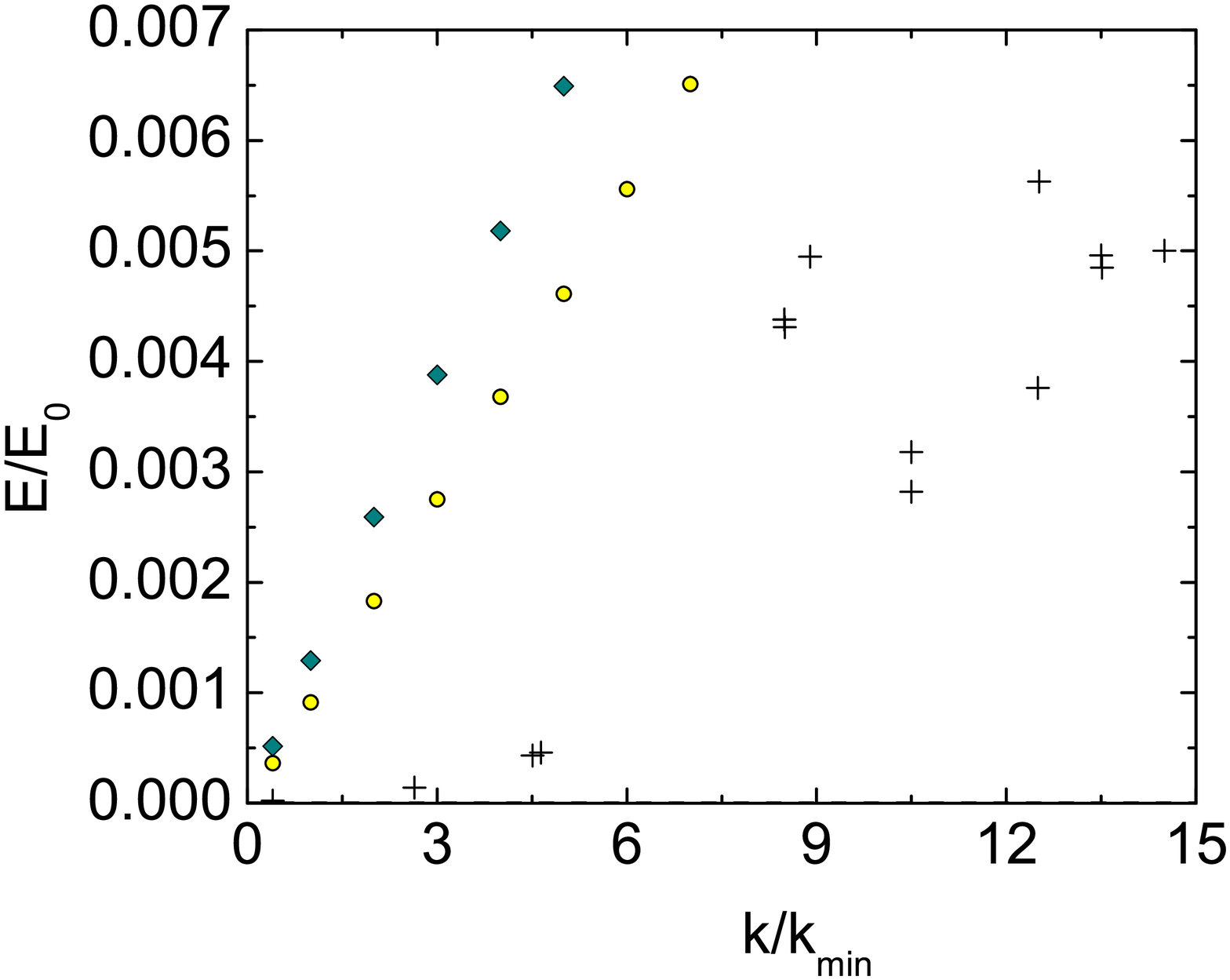}
\hfill
\includegraphics[width=0.47\textwidth]{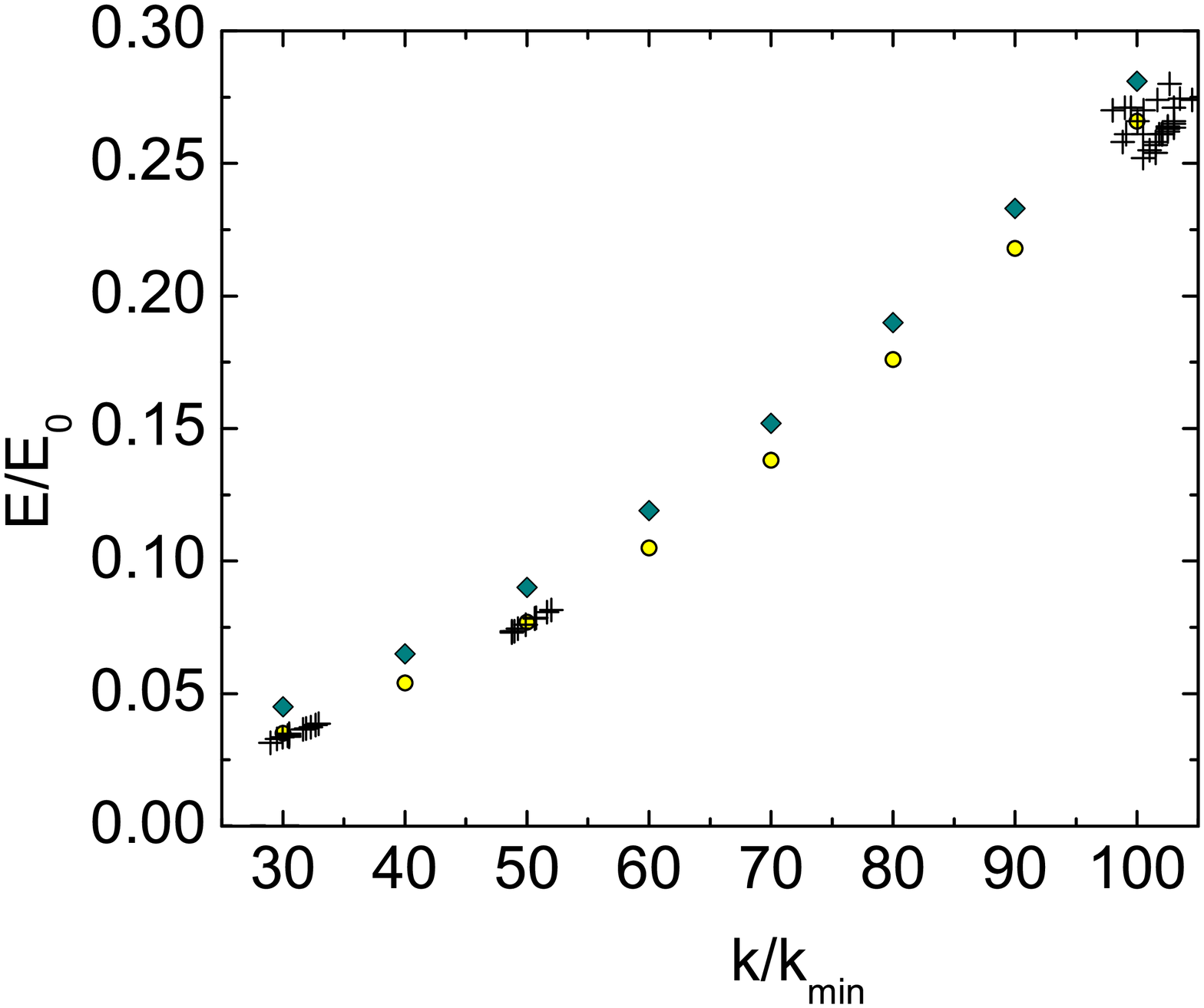}
\\
\parbox[t]{0.47\textwidth}{
\caption{ Dispersion laws: Bogolyubov (\ref{2}) (rhombs); curve (\ref{5}),
(\ref{31}) with $q=1/2$ (circles); and numerical solutions of system
(\ref{51}) and (\ref{53}) at $z=-1$ (crosses). Here, $k_{min}=\pi/L$,
$E_{0}=\hbar^{2}k_{m}^{2}/2m$, and $k_{m}=2\pi/\bar{R}$. Hundred ($N=100$)
He$^{4}$ atoms with interatomic potential (\ref{33}) were considered;
$U_{0}=0.1~\mathrm{K}$ and $a=0.1\bar{R}$.
 \label{fig2}} } \hfill
\parbox[t]{0.47\textwidth}{
\caption{ The same as in Fig.~\ref{fig2}, but for large $k$'s.   \label{fig3}} }
\end{figure}

The frequencies are determined as follows. System (\ref{51}) looks like%
\begin{equation}
\sum\limits_{p=1,2,\ldots}a_{p}A_{jp}=0,\ j=1,2,\ldots. \label{54}%
\end{equation}
The quantity $\hbar^{2}\omega^{2}$ in the coefficients $A_{jp}$ are taken in
form (\ref{30}) with $k_{0}=\pi j_{0}/L$ and $j_{0}=100$, and $q^{eff}$ is
smoothly varied from $-100$ to 1000. At some $q^{eff}$'s, the absolute value
of $A_{jp}$-matrix determinant drastically decreases (approximately by two
orders of magnitude), and those $q^{eff}$-values are characteristic
frequencies, the roots $\hbar^{2}\omega^{2}$ of equation $\mathrm{det}%
~A_{jp}=0$. Dispersion law (\ref{31}) is found with the help of relation
(\ref{32}). A numerical calculation was made for $N=100$ and $r_{1}$ from
Fig.~\ref{fig1}. A $2N\times2N$ matrix was used to represent $A_{jp}$ (the
increase of matrix size affected the results very weakly). The interatomic
potential was simulated by formula (\ref{33}) with $U_{0}=0.1~\mathrm{K}$ and
$a=0.1\bar{R}$. In Figs.~\ref{fig2} and \ref{fig3}, the frequency values are
depicted. The coefficient $q$ in dispersion law (\ref{31}) is close to 1/2
for relatively large $k$'s (for the Bogolyubov solution, $q=1$). For small
$k$'s, the magnitude of $q$ is smaller; it
differs for different
roots and depends on $U_{0}$ and $a/\bar{R}$. Besides, $E$ weakly depends on
$\gamma$ and $r_{1}$. Its dependence on $r_{1}$ is associated with the error
arising owing to the account of the finite number of summands in expansion
(\ref{47}). The dependence on $\gamma$ is conspicuous at small $k$'s, when the
ratio between the fractional, $\pi\gamma/L$, and integer, $\pi j/L$, parts of
$k$ is not small.

Note that the network of smallest roots $q^{eff}$ is extremely dense at $j_{0}%
\gg1$, so that $q^{eff}$ must be varied with a very short increment for not to
miss any root. There are no lost roots if the number of roots equals the
number of rows in our square matrix.

As the initial $\gamma$-value varies, the theoretical $\gamma$ obtained from
Eq.~(\ref{53}) oscillates with various amplitudes of an order of 1. Often, the
period of oscillations is very short, e.g., 10$^{-6}$ or even 10$^{-9}$. This
is possible, because the system of equations is large and strongly nonlinear
with respect to $\gamma$. The majority of solutions lie on such small
oscillations. The periods of some oscillations are extremely short, so that
the corresponding solutions can be easily overlooked. They must be found by
feeling, and the procedure of finding them takes a lot of time. Unfortunately,
the method does not guarantee that all the solutions have been found. One
cannot even say how many solutions exist for every frequency root.

The solutions that we managed to find are depicted in Figs.~\ref{fig2} and
\ref{fig3} by crosses. They determine a dispersion curve lying below the
Bogolyubov one. Moreover, we did not obtain a curve but a strip consisting of
a good many points (the M-strip, a derivative of \textquotedblleft
many\textquotedblright). Why did we obtain a strip rather than a curve? This
can be 1)~a result of zero BCs, 2)~owing to the coordinate-momentum
uncertainty relation for the quasiparticle, and 3)~the strong spread of points
can be associated with the fact that the wave packet does not correspond to
the solution exactly (instead of one maximum, the packet usually includes two
close maxima, see section \ref{sec5}).

In order to plot the dispersion curve, one should know the $E$- and $k$-values
for every point. The energy can be found unambiguously from the equations,
whereas $k=k^{wp}$ is determined from the wave-packet structure, inexactly if
$k$ is small (see section \ref{sec5}). The energy $E$ is  low at small
$k$ and is also found inexactly  if the determination error $\triangle E\sim
E$. Therefore, the nonlinearity of
the dispersion curve at small $k$ (see
Fig.~\ref{fig2}) are connected, most likely, with the calculation error, so
that if the calculation were exact, we would obtain the linear law of the type
$E=ck$ with $q\approx1/2$. (According to general theorems \cite{yukalov2011}, phonons in the
superfluid Bose liquid originates from a spontaneous violation of gauge
symmetry and have to possess the asymptotic $E(k\rightarrow0)=ck$.
However, the theorems were proved for periodic BCs, and we
do not know whether they remain valid for zero ones.)

Let us return to the general procedure. We solved the linear system of 200
equations (\ref{51}). For a given $\gamma$, it has 200 solutions for
$\omega^{2}$. Plotting the dependences of all 200 frequencies on the initial
$\gamma$ in the interval $\gamma\in\left]  -1,1\right]  $, we obtain a network
of 200 lines. Each of them corresponds to a frequency with a certain number.
In this network, the lines sometimes intersect each other and break. In the
latter case, a new line emerges elsewhere. For every line, there are several
points where the initial and theoretical $\gamma$'s coincide. Just those
points are the system energy levels. We found solutions only for $z=-1$. If
$z=1$, the equations are similar, and so must be their solutions. As a result,
the number of levels in the M-strip will be approximately doubled.

For the first (lowest) energy level, we found 2 solutions (we designate this
as $1_{2}$), and for the next levels $2_{2}$, $8_{3}$, $9_{9}$,
$29_{5}$, $30_{8}$, $49_{7}$, $50_{3}$, $99_{16}$, and $100_{13}$. They are
shown in Figs.~\ref{fig2} and \ref{fig3}. For every frequency solution, we
solved Eq.~(\ref{51}) to find the coefficients $a_{p}$ for wave packet (2b)
(Eq.~(\ref{2b-1})). As a rule, the packet included two closely located maxima
(see Fig.~\ref{fig4}), over which we  determined by eye the $p$-index
value corresponding to the packet center and found the packet wave vector $\tilde
{k}_{p}=\pi(p+\gamma(p))/L$. Some packets
had 1 or 3 maxima. Those $\tilde{k}_{p}$'s were substituted as $k$ into the
dispersion law $E(k)$.

The main result of this section consists in that wave packets (2a) and (2b)
are characterized by a new dispersion law (\ref{31}) with $q\approx1/2$.

\section{Solution structure}

\label{sec5}

In sections \ref{sec2} and \ref{sec3}, we found that the Bogolyubov dispersion
law corresponds to solutions~(1a) with $\gamma=\pm1/2$ and (1b) with
$\gamma=0$. The packet wave vector $k^{wp}=2\pi(l_{0}+\gamma)/L>0$. The
structure of packet~(1a) was found by substituting the solution for $\omega$
with the $j$-th number into Eq.~(\ref{28}) and solving the latter with respect
to the coefficients $a_{2l}$. The coefficient $a_{2l}$ with $l=j$ turned out
the largest one, the neighbor coefficients $a_{2l}$ (with $l=j\pm1$) were less
by 2 to 3 orders of magnitude, and the next coefficients were even less. It
means that the packet is strongly localized in the $k$-space, as well as
packet~(1b) \cite{zero-gas2}. Therefore, it is easy to find the value
$k=k^{wp}$ for the dispersion law $E(k)$: we should use in the formula $k^{wp}=2\pi(l_{0}+\gamma)/L$
the number $l_{0}=j$ of the largest $a_{2j}$.

For the M-mode (Figs.~\ref{fig2} and \ref{fig3}), the situation is much more
complicated. The basic issues are: What is the nature of this solution? Is it
a new oscillatory mode or a superposition of several Bogolyubov modes? The
formula for the wave packet reads
\begin{align}
&  \tilde{n}(x,k^{wp})=2\sum\limits_{l=1,2,\ldots}a_{2l}(l_{0})\sin
{[\pi(2l+\gamma)x/L]}\nonumber\\
&  +2\sum\limits_{j=0,1,2,\ldots}a_{2j+1}(l_{0})\sin{[\pi(2j+1+\gamma
)x/L]}\nonumber\\
&  =\sum\limits_{l=\pm1,\ldots}a_{2l}e^{i\pi(2l+\gamma(2l))x/L} \nonumber\\
&+\sum
\limits_{j=0,\pm1,\ldots}a_{2j+1}e^{i\pi(2j+1+\gamma(2j+1))x/L}.\label{6-10}%
\end{align}
It should be appended by Eqs.~(\ref{42}), (\ref{43}), $z=-1$, and by BC
(\ref{53}). Harmonics with odd numbers can be expanded in harmonics with even
numbers, i.e. in the Fourier series,
\[
e^{i\pi(2j+1)x/L}=\frac{2}{\pi}\sum\limits_{p}\frac{\sin{[\pi(2p-2j-1)/2]}%
}{2p-2j-1}e^{i\pi2px/L},
\]
where $p$'s are integers. Then, wave packet (\ref{6-10}) looks like
\begin{align}
&  \tilde{n}=2\sum\limits_{l=0,1,2,\ldots}\tilde{a}_{2l}(l_{0})\sin
{[\pi(2l+\gamma)x/L]}\nonumber\\
&  +2\sum\limits_{l=1,2,\ldots}\tilde{c}_{2l}(l_{0})\sin{[\pi(2l-\gamma
)x/L]},\label{6-3}%
\end{align}
i.e. it is a sum of two packets with \textquotedblleft even\textquotedblright%
\ harmonics, where $\gamma$ is left as a quantum number. For every
\textquotedblleft even\textquotedblright\ packet, the Bogolyubov dispersion
law is valid, but the zero BC is not obeyed. If the zero BC has been satisfied
for each of those packets at any frequency, packet~(\ref{6-10}) would have
been a sum of two or more Bogolyubov packets. However, the zero BC is obeyed
only for the whole sum (\ref{6-10}); therefore, the M-mode is not reduced to a
sum of several solutions of the Bogolyubov type. Hence, this is a new
solution. The fact that it is constructed on the basis of two basis sets makes
the solution less clear and a little debatable. However, it seems that the
packet center $k_{wp}$ and the dispersion law $E(k)$ can be indicated for the
new mode, at least if $k_{wp}$'s are not small.

\begin{figure}[ht]
\centerline{\includegraphics[width=85mm]{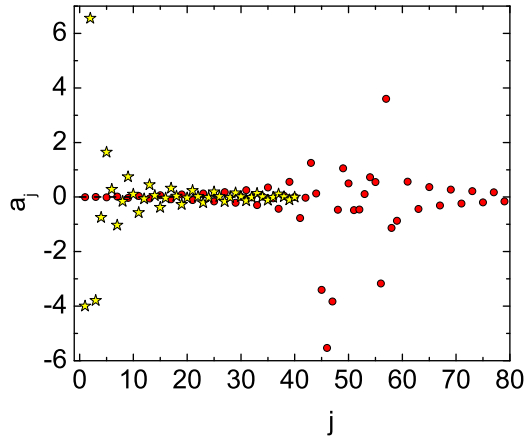}
}
\caption{ The set of coefficients $a_{j}$ for wave packet (\ref{6-10}). The
solutions of system (\ref{51}) and (\ref{53}) with $z=-1$ are
considered for the frequency No.~1 with $E\approx0.000142E_{0}$,
$\gamma\approx0.6435$, and $k^{wp}\approx2.64\pi/L$ (stars), and for frequency
No.~50 with $E\approx0.0782E_{0}$, $\gamma\approx-0.6065$, and $k^{wp}%
=\pi(j^{wp}+\gamma)/L\approx50.64\pi/L$ (circles; the absolute values of
$a_{j}$'s are decreased by a factor of 30). Hundred ($N=100$) He$^{4}$ atoms
with interatomic potential (\ref{33}) were considered; $U_{0}=0.1~\mathrm{K}$
and $a=0.1\bar{R}$.
 \label{fig4}} \end{figure}
Let us consider the structure of packet (\ref{6-10}). For every $\omega$, the
coefficients $a_{j}$ are determined from system (\ref{51}) after substituting
the solution for $\omega$ into it.  One can see
from the results (Fig.~\ref{fig4})
that the packet is not strongly localized in the $k$-space.
However, the $k^{wp}$-value can be approximately calculated. At large $k^{wp}%
$, the wave packet includes two close maxima, so that $k^{wp}$ can be
evaluated as a half-sum of maximum $k$-values (if either of the maxima is
larger, $k^{wp}$ is shifted proportionally). The error obtained for $k^{wp}$
in this case is small, and the value of $E$ for the packet is uniquely
determined from the system of equations. Therefore, at large $k^{wp}$,
M-packet (\ref{6-10}) rather adequately describes the quasiparticle. At small
$k^{wp}$, the wave packet has one or two, not very narrow, maxima, the
$k$-values of which slightly exceed the increment step of $k$. In this case,
the value of $k^{wp}$ is determined unreliably. Most likely, the dispersion
curve (Fig.~\ref{fig2}) has a \textquotedblleft dip\textquotedblright\,
because small $k^{wp}$'s were determined
inaccurately. Perhaps, there may exist another representation of packet
(\ref{6-10}) as a sum of harmonics, for which the wave packet has a single
sharp maximum at small $k^{wp}$. If we associate $k^{wp}$ with this maximum,
the dispersion law should probably be of the linear type, $E=ck$. However, we did
not find such a representation.
\begin{figure}[ht]
\centerline{\includegraphics[width=85mm]{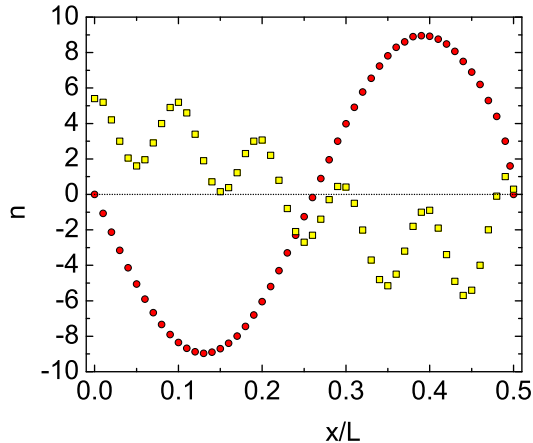}
}
\caption{Dependence $\tilde{n}(x)$ for new mode (\ref{6-10}) for one of the
solutions for the 9-th (circles) and 30-th (squares) frequencies. For the 9-th
frequency, $E\approx0.005E_{0}$, $\gamma\approx-0.1$, and $k^{wp}\approx
8.9\pi/L$; and for the 30-th one, $E\approx0.036E_{0}$, $\gamma\approx0.64$,
and $k^{wp}\approx31.64\pi/L$L. Hundred ($N=100$) He$^{4}$ atoms with
interatomic potential (\ref{33}) were considered; $U_{0}=0.1~\mathrm{K}$ and
$a=0.1\bar{R}$. For illustrative purposes, the curve for the 30-th frequency
is stretched tenfold along the $x$-axis, so that $x=0.5L$ in the figure
corresponds to the real $x=0.5L$, and $x=0$ in the figure to real $x=0.45L$.
  \label{fig5}}
\end{figure}

It is of interest that, in a more exact approach \cite{zero-liquid}, the
solution for the phonon is so constructed that the phonon is described by
a single harmonic rather than a wave packet. As a result, $k$ is determined
unambiguously. In this case, $E(k\rightarrow0)=ck$.

Figure~\ref{fig5} illustrates solution (\ref{6-10}) in the $x$-space
(for two arbitrary $E$-roots). The curve composed by squares looks rather
strange. This density distribution is probably unstable. Notice that, at small
$k^{wp}=\pi(l_{0}+\gamma)/L$, the number of zeros for the function $\tilde
{n}(x)$ is considerably less than that for $\sin{k^{wp}x}$, with those two
numbers getting closer as ${k^{wp}}$ grows.

\section{Thermodynamics}

\label{sec6}

According to the analysis given above, the following points correspond to the
Bogolyubov curve: one $\omega$ for every $k=2l\pi/L$, two very close $\omega
$'s for every $k=(2l+1)\pi/L$ ($l=1,2,\ldots$), and one $\omega$ for $k=\pi
/L$. We may average over the nearest levels and consider that any interval
$\triangle k=\pi/L$ includes $\alpha(k)=1.5$~points. The free energy of the
system equals \cite{land5}%
\begin{equation}
F=-k_{B}T\ln\sum\limits_{n}e^{-E_{n}/k_{B}T}. \label{5-1}%
\end{equation}
In our case,
\begin{equation}
F=-k_{B}T\ln{\left(  \sum\limits_{k=\pi j/L>0}\alpha(k)e^{-E_{k}/k_{B}%
T}\right)  }. \label{5-2}%
\end{equation}
For the Bogolyubov curve, $\alpha(k)$ does not depend on $k$ and leads only to
the summand $-k_{B}T\ln\alpha$. It is unobservable, because it changes the
system entropy $S=-\partial F/\partial T$ but not the total energy $E=F+TS$
and the heat capacity $C=T\partial S/\partial T$.

The M-strip gives its contribution to $F$. According to available data,
$\alpha(k)\sim5$ for frequencies with  numbers $1$-$5$, and $\alpha(k)\sim10$ for
frequencies with numbers $\gtrsim10$. Moreover, $\alpha$ chaotically changes
when passing to next levels. No evident growth or decrease of $\alpha$ was
observed as $k$ grew.

\section{Comparison with experiment}

\label{sec7}

We found two possible dispersion laws for the uniform 1D Bose gas in a box:
the Bogolyubov law and a new one (\ref{5}) with $q=1/2$. A similar
problem was solved for the $f$-dimensional case using a more exact method
\cite{zero-liquid}, and dispersion law (\ref{5}) with $q=2^{-f}$ was obtained.
Unlike the GP method, the approach \cite{zero-liquid} catches the
structure of the system ground state. In this approach,
different ground states (with different energies $E_{0}$) correspond to the
Bogolyubov and new dispersion laws. Therefore, the system should be so ordered
that either all phonons obey the Bogolyubov dispersion law or they have
dispersion law (\ref{5}) with $q=2^{-f}$. In the 2D and 3D cases, $E_{0}$ is
lower for the new solution \cite{zero-liquid}; therefore, it has to be
realized in the Nature, whereas the Bogolyubov ordering should be unstable.

How can all that be verified experimentally? The dispersion law should be
measured for the uniform Bose gas, with boundaries and cyclic, and the
results should be compared. A cyclic system of any dimensionality is characterized by the
Bogolyubov dispersion law only. As far as we
know, uniform 1D systems have not been created yet. A three-dimensional system
cannot be made closed. However, such an experiment is quite possible for
2D films; and this is probably the main way of verification
\cite{zero-liquid}.

A huge number of experiments are carried out now with gases in traps, but they
seem not to be useful in verifying the effect. The reason is as follows. In
the case of uniform system, when changing from periodic BCs to zero ones, the
second dispersion law manifests itself owing to the interaction between the
harmonics in the potential expansion and the harmonics in the expansion of
density oscillations in the integral in the Gross equation. Gas in a
trap is a cloud, which is dense near the center. At a distance of about the
Thomas-Fermi length from the center, the concentration becomes low and rapidly
decreases further. The cloud is in a vessel, the size of which is much larger
than the Thomas-Fermi length; therefore, the gas atoms are practically absent
near the vessel walls. For this system, oscillatory functions are
characteristic ones. If the system becomes closed in the region of boundaries,
oscillatory functions change only in this region. However, both the
concentration and the low-order oscillatory functions are small there;
therefore, the integral in the Gross equation must be almost identical under
zero and periodic BCs. Accordingly, the dispersion laws must also be close to
each other. If the system is uniform, the characteristic functions are plane
waves (not small in the \textit{whole} space), and the change from periodic to zero BCs
manifests itself in Eq.~(\ref{20}) as the increment modification, $\triangle
k=2\pi/L\rightarrow\pi/L$, and a factor of 1/2 before $\nu(k)$. As a result,
the integral in the Gross equation brings about different solutions for
periodic and zero BCs.

For gases in the trap, the integral in the Gross equation does not feel the
difference between the topologies of closed and open systems. Therefore, it
can be calculated approximately, assuming the potential to be point-like and
thus changing to the GP equation. This is an ordinary way. The GP equation
leads to the Bogolyubov formula for the dispersion law in a uniform system.
This circumstance favors the application of this formula rather than
Eq.~(\ref{5}) as the basic dispersion law in the local density approximation
(LDA). In the work by Stringari \cite{stringari1996}, the low-lying 3D levels
were calculated from the GP equation, and they agreed with the experiment
\cite{cornell1996,ketterle1996,ketterle2000}. For a 1D gas in the
trap, experimentally measured lower levels \cite{esslinger2003,nagerl2009}
approximately agree with theoretical ones
\cite{stringari2002,santos2003,superTG}. For the linear and square-law
regions in the dispersion curve in the 3D geometry, the theory
(\cite{zaremba1998,tozzo2003} and LDA) approximately agrees with the
experiment for $^{23}$Na \cite{ketterleE1,ketterle1999,ketterleE2} and $^{87}%
$Rb \cite{steinhauer2002,steinhauer2003,steinhauer2012} atoms (see also review
\cite{ozeri2005}).

Notice that, in the often used LDA, the dispersion law for a localized wave
packet in a non-uniform medium is assumed \cite{zambelli2000} to be the same
as for a packet in an infinite uniform environment of the same density.
However, the localized packet is a superposition of nonlocal modes; therefore,
the dispersion law for the packet can turn out a nonlocal property. For today,
there is no clear evidence that this is so, but the rigorous substantiation of
LDA is also absent. The accuracy of LDA  can be elucidated by
constructing a wave packet on the basis of exact solutions for nonlocal modes
of non-uniform system.

\section{Conclusions}

In this work, we tried to find a dispersion law for the one-dimensional
\textit{uniform} Bose gas under zero boundary conditions. We proceeded from
the Gross equation (\ref{4}) making allowance for the non-point character of
interaction. Two solutions were found: the known Bogolyubov dispersion curve
(which is reproduced with a high accuracy) and a new curve $E(k)$ that lies
below the Bogolyubov one. The new curve can be reliably determined at large
$k$'s and not so reliably at small $k$'s. It agrees well with a solution for
$E(k)$ found  by a different method \cite{zero-liquid}
for a one-, two-, and three-dimensional systems. The stability of solutions was not analyzed. Which
of two dispersion laws is realized in experiment can be elucidated by studying
homogeneous rarefied He~II films \cite{zero-liquid}.

An actual non-point potential is replaced by the point-like one, because
the former is very complicated and is not known precisely, whereas the latter
simplifies the equations very much. As far as we understand, this replacement
is justified for the description of atomic scattering. However, for studying
the collective properties of Bose gas at $T\rightarrow0$, it is not justified,
since the new solution becomes lost. Collective oscillations and Fourier
components of potential are modulated by the walls \cite{zero-liquid}, and
they mutually interact in the integral in the Gross equation. If the non-point
character of interaction is taken into account, the Fourier components
strongly change, and a new solution for collective modes emerges.

The problem of an influence of boundaries is interesting but not simple. To make this
issue clear, unbiased researches using various theoretical methods and special
experiments are required. We hope that this work will be useful.




\begin{thebibliography}{99}                                                                                               %


\bibitem {girardeau1960}M. Girardeau, J. Math. Phys. (N.Y.) \textbf{1}, 516 (1960).

\bibitem {lieb1963}E.H. Lieb, W. Liniger, Phys. Rev. \textbf{130}, 1605
(1963); E.H. Lieb, Phys. Rev. \textbf{130}, 1616 (1963).

\bibitem {gaudin}M. Gaudin, Phys. Rev. A \textbf{4}, 386 (1971).

\bibitem {stringari2002}C. Menotti and S.~Stringari, Phys. Rev. A \textbf{66},
043610 (2002).

\bibitem {santos2003}P. Pedri and L.~Santos, Phys. Rev. Lett. \textbf{91},
110401 (2003).

\bibitem {superTG}G.E. Astrakharchik, J.~Boronat, J.~Casulleras, and
S.~Giorgini, Phys. Rev. Lett. \textbf{95}, 190407 (2005).

\bibitem {esslinger2003}H. Moritz, T.~St\"{o}ferle, M.~K\"{o}hl, and
T.~Esslinger, Phys. Rev. Lett. \textbf{91}, 250402 (2003).

\bibitem {nagerl2009}E. Haller, M.~Gustavsson, M.J.~Mark, J.G.~Danzl, R.~Hart,
G.~Pupillo, and H.-C.~N\"{a}gerl, Science \textbf{325}, 1224 (2009).

\bibitem {pit1961}L.P. Pitaevskii, Sov. Phys. JETP \textbf{13}, 451 (1961).

\bibitem {gross1963}E.P. Gross, J. Math. Phys. \textbf{4}, 195 (1963).

\bibitem {gross1957}E.P. Gross, Phys. Rev. \textbf{106}, 161 (1957).

\bibitem {zero-liquid}M.D. Tomchenko, Ukr. J. Phys. \textbf{59}, 123 (2014); arXiv:cond-mat/1201.1845.

\bibitem {bog1947}N.N. Bogoliubov, J. Phys. USSR \textbf{11}, 23 (1947).

\bibitem {bz1955}N.N. Bogolyubov and D.N. Zubarev, Sov. Phys. JETP \textbf{1},
83 (1956).

\bibitem {zero-gas2}M. Tomchenko, arXiv:cond-mat/1211.1723.

\bibitem {druten1997}N.J. van Druten and W.~Ketterle, Phys. Rev. Lett.
\textbf{79}, 549 (1997).

\bibitem {girardeau2001}M.D. Girardeau, E.M.~Wright, and J.M.~Triscari, Phys.
Rev. A \textbf{63}, 033601 (2001).

\bibitem {shlyap2004}D.S. Petrov, D.M. Gangardt, and G.V. Shlyapnikov, J.
Phys. IV France \textbf{116}, 5 (2004); arXiv:cond-mat/0409230.

\bibitem {pethick2008}C.J.~Pethick, H.~Smith, \textit{Bose-Einstein
Condensation In Dilute Gases} (Cambridge University Press, New York, 2008).

\bibitem {ryady}M.D. Tomchenko, arXiv:cond-mat/1403.8014.

\bibitem {yukalov2011}V.I. Yukalov, Phys. Part. Nucl. \textbf{42}, 460 (2011).

\bibitem {land5}L.D.~Landau and E.M.~Lifshitz, \textit{Statistical Physics},
3rd ed., Part 1, (Pergamon Press, Oxford, 1980; Nauka, Moscow, 1976).

\bibitem {stringari1996}S. Stringari, Phys. Rev. Lett. \textbf{77}, 2360 (1996).

\bibitem {cornell1996}D.S.~Jin, J.R.~Ensher, M.R.~Matthews, C.E.~Wieman, and
E.A.~Cornell, Phys. Rev. Lett. \textbf{77} 420 (1996).

\bibitem {ketterle1996}M.-O.~Mewes, M.R.~Andrews, N.J.~van~Druten, D.M.~Kurn,
D.S.~Durfee, C.G.~Townsend and W.~Ketterle, Phys. Rev. Lett. \textbf{77} 988 (1996).

\bibitem {ketterle2000}R.~Onofrio, D.S.~Durfee, C.~Raman, M.~K\"{o}hl,
C.E.~Kuklewicz, and W.~Ketterle, Phys. Rev. Lett. \textbf{84} 810 (2000).

\bibitem {zaremba1998}E. Zaremba, Phys. Rev. A \textbf{57}, 518 (1998).

\bibitem {tozzo2003}C.~Tozzo and F.~Dalfovo, New J. Phys. \textbf{5}, 54 (2003).

\bibitem {ketterleE1}M.R. Andrews, D.M.~Kurn, H.-J. Miesner, D.S.~Durfee,
C.G.~Townsend, S.~Inouye, and W.~Ketterle, Phys. Rev. Lett. \textbf{79}, 553
(1997); M.R.~Andrews, D.M.~Stamper-Kurn, H.-J.~Miesner, D.S.~Durfee,
C.G.~Townsend, S.~Inouye, and W.~Ketterle, Phys. Rev. Lett. \textbf{80}, 2967
(E) (1998).

\bibitem {ketterle1999}D.M.~Stamper-Kurn, A.P.~Chikkatur, A.~G\"{o}rlitz,
S.~Inouye, S.~Gupta, D.E.~Pritchard, and W.~Ketterle, Phys. Rev. Lett.
\textbf{83}, 2876 (1999).

\bibitem {ketterleE2}J. Stenger, S.~Inouye, A.P.~Chikkatur, D.M.~Stamper-Kurn,
D.E.~Pritchard, and W.~Ketterle, Phys. Rev. Lett. \textbf{82}, 4569 (1999).

\bibitem {steinhauer2002}J. Steinhauer, R. Ozeri, N. Katz, and N. Davidson,
Phys. Rev. Lett. \textbf{88}, 120407 (2002).

\bibitem {steinhauer2003}J. Steinhauer, N. Katz, R. Ozeri, N. Davidson,
C.~Tozzo, and F.~Dalfovo, Phys. Rev. Lett. \textbf{90}, 060404 (2003).

\bibitem {steinhauer2012}I. Shammass, S. Rinott, A. Berkovitz, R. Schley, and
J. Steinhauer, Phys. Rev. Lett. \textbf{109}, 195301 (2012).

\bibitem {ozeri2005}R. Ozeri, N.~Katz, J.~Steinhauer, and N.~Davidson, Rev.
Mod. Phys. \textbf{77},187 (2005).

\bibitem {zambelli2000}F. Zambelli, L.~Pitaevskii, D.M.~Stamper-Kurn, and
S.~Stringari, Phys. Rev. A \textbf{61}, 063608 (2000).
\end{thebibliography}
\end{document}